\documentstyle[aps,prb]{revtex} 
\begin{document}


\twocolumn[\hsize\textwidth\columnwidth\hsize\csname
@twocolumnfalse\endcsname

\title{Self-energy of image states on copper surfaces}
\author{I. Sarria$^1$, J. Osma$^2$, E. V. Chulkov$^{2,3}$, J. M. Pitarke$^{1,3}$, and
P. M. Echenique$^{2,3}$}
\address{$^1$Materia Kondentsatuaren Fisika Saila, Zientzi Fakultatea,
Euskal Herriko Unibertsitatea,\\ 
644 Posta Kutxatila, 48080 Bilbo, Basque Country, Spain\\
$^2$Materialen Fisika Saila, Kimika Fakultatea, Euskal Herriko
Unibertsitatea,\\  1072 Posta kutxatila, 20080 Donostia, Basque Country,
Spain\\
$^3$Donostia International Physics Center (DIPC) and Centro Mixto CSIC-UPV/EHU,
Spain}

\date\today

\maketitle

\begin{abstract}
We report extensive calculations of the imaginary part of the electron
self-energy in the vicinity of the (100) and (111) surfaces of Cu. The
quasiparticle self-energy is computed by going beyond a free-electron
description of the metal surface, either within the GW approximation of
many-body theory or with inclusion, within the GW$\Gamma$
approximation, of short-range exchange-correlation effects. Calculations of the
decay rate of the first three image states on Cu(100) and the first image state
on Cu(111) are also reported, and the impact of both band structure and many-body
effects on the electron relaxation process is discussed.
\end{abstract}
\pacs{PACS numbers: 71.45.Gm, 73.20.At, 78.47.+p, 79.60.Bm}
]


\section{Introduction}
It is well-known that an electron outside a metal surface experiences an
effective potential with the asymptotic image form $V_{im}=-1/4(z-z_0)$, $z_0$
representing the image-plane position. If the bulk band structure projected
into the surface presents a band gap near the vacuum level ($E_v=0$), an
electron with energy $E<0$ can be trapped in the potential-well created on the
vacuum side of the surface by the gap and the image potential. The long range
character of the image potential gives rise to a series of unoccupied
Rydberg-like states, which converge towards the vacuum
level.\cite{Echenique1,Osgood} These so-called image states were first detected
experimentally\cite{Dose,Himpsel} by the techniques of inverse
photoemission,\cite{Pendry,Smith1} and the first high-resolution measurements of
image states were made by the use of two-photon photoemission
(2PPE).\cite{Giesen,Padowitz,Fauster0,Fauster,Harris}

Image-potential states are localized in the vacuum region of the surface.
Hence, they are almost decoupled from bulk states and are expected to
have much longer lifetimes than bulk excitations. Lifetimes of image states
had been determined from their spectral width in 2PPE spectroscopy, and
the advent of time-resolved 2PPE (TR-TPPE) has enabled a direct measurement
of image-state lifetimes on the (100) and (111)
surfaces of silver\cite{Fuji1,Fuji2,Fuji3} and
copper.\cite{Hertel1,Hertel2,Hertel3,Hofer1,Hofer2}

Calculations of image-state lifetimes were reported in
Refs.\onlinecite{Echenique3} and\onlinecite{Echenique4}, within a many-body
free-electron description of the metal surface and with the use of simplified
models to approximate both initial and final electronic states and, also, the
screened Coulomb interaction. Later on, the role that crystal-induced surface
states, not present within a free-electron description of the solid, play in
the decay of image states was investigated.\cite{Gao1,Osma}
Self-consistent many-body calculations of image-state lifetimes that go beyond a
free-electron description of the metal surface have been carried out only very
recently on copper\cite{Chulkov1} and lithium\cite{Chulkov2} surfaces.

In this paper we report extensive calculations of the imaginary part of the
image-electron self-energy in the vicinity of the (100) and (111) surfaces
of copper, which we compute within the GW approximation of many-body
theory.\cite{Hedin} Single-particle wave functions are obtained by solving the
Schr\"odinger equation with a realistic one-dimensional model potential, and the
screened interaction is evaluated within the random-phase approximation
(RPA).\cite{Pines} Then, we present the results of calculations of the lifetime
of the first three image states on Cu(100) and the first image state on Cu(111).
These calculations are carried out within a GW-RPA description of the
self-energy and, also, with use of the so-called GW${\rm\Gamma}$
approximation,\cite{Mahan1,Mahan2} which includes short-range
exchange-correlation (XC) effects not present in the GW-RPA. We also consider
simplified models for the description of both the electronic wave functions and
the screened interaction, we discuss the impact of band structure and many-body
effects on image-state lifetimes, and account for the potential variation
parallel to the surface through the introduction of the effective mass. We
present systematic investigations of the role that all quantities involved in
the evaluation of the electron self-energy play in the coupling of image
states with the solid. We find that a detailed description of thes quantities is
of crucial importance in the understanding of the origin and magnitude of decay
rates of image states. Finally, differences between decay rates of image states
on the (100) and (111) surfaces of Cu are investigated, and we focus our
attention on the role that the various available bulk and surface states play in
the electron relaxation process.

Unless otherwise is stated, atomic units  are used throughout, i. e.,
$e^2=\hbar=m_e=1$. 

\section {Theory}

Take an inhomogeneous electron system. The damping rate of an excited electron
in the state $\psi_0({\bf r})$ with energy $E_0$ is obtained as the projection of
the imaginary part of the electron self-energy, $\Sigma({\bf r},{\bf r}',E_0)$,
over the state itself:
\begin{equation}\label{eq1}
\tau^{-1}=-{2}\int{\rm d}{{\bf r}}\int{\rm d}{{\bf
r}'}\,\psi_{0}^*({\bf r})\,{\rm Im}\,\Sigma({\bf r},{\bf r}';E_0)\,
\psi_{0}({\bf r}').
\end{equation}

We consider a periodic-film model of the solid. The film is taken to be
translationally invariant in the plane of the surface, which is assumed to be
normal to the $z$ axis, and departure of motion along the surface from
free-electron behaviour is accounted through the introduction of the effective
mass. Hence, single-particle wave functions are taken to be of the form
\begin{equation}\label{3} 
{\psi}_0({\bf r})={1\over\sqrt A}\,\phi_0(z)\,e^{{\rm i}\,{\bf
k}_\parallel\cdot {\bf r}_\parallel},
\end{equation}
with energies
\begin{equation}\label{4}
E_0=\varepsilon_0+{{\bf k}_\parallel^2\over 2\,m_0}.
\end{equation}
The wave function $\phi_0(z)$ and energy $\varepsilon_0$ describe motion
normal to the surface, ${\bf k}_\parallel$ is a wave vector parallel to the
surface, $A$ is the normalization area, and $m_0$ is the effective mass.
Introduction of Eq. (\ref{3}) into Eq. (\ref{eq1}) yields the following
expression for the damping rate: 
\begin{equation}\label{5}
\tau^{-1}=-2\int{\rm d}z\int{\rm d}z'\,\phi_0^*(z)
\,{\rm Im}\,\Sigma(z,z';{\bf k_\parallel},E_0)\,\phi_0(z'),
\end{equation}
where $\Sigma(z,z';{\bf k_\parallel},E_0)$ represents the
two-dimensional Fourier transform of the electron self-energy.

In the so-called $GW$ approximation, one considers only the first-order term in
a series expansion of the self-energy in terms of the screened interaction,
\begin{eqnarray}\label{eq13}
\Sigma(z,z';{\bf k_\parallel},E_0)=&&
\int_{-\infty}^{\infty}{{\rm d}E\over 2\pi}\int{{\rm d}{\bf
q}_\parallel\over(2\pi)^2}\,W(z,z';{\bf q}_\parallel,E)\cr\cr
&&\times\,{\rm i}\,G(z,z';{\bf k_\parallel}-{\bf
q_\parallel},E_0-E),
\end{eqnarray}
and after replacing the Green function $G(z,z';{\bf k_\parallel},\omega)$
by that of non-interacting electrons, one finds
\begin{eqnarray}\label{6}
{\rm Im}\,\Sigma(&&z,z';{\bf k}_\parallel,E_0)=\sum_f
\,\phi_f^*(z')\,\phi_f(z)\cr\cr 
&&\times\int{{\rm d}{\bf q}_\parallel\over(2\pi)^2}{\rm Im}\,W(z,z';{\bf
q}_\parallel,E_0-E_f),
\end{eqnarray}
where
\begin{equation}\label{4}
E_f=\varepsilon_f+{({\bf k}_\parallel+{\bf q}_\parallel)^2\over 2\,m_f}.
\end{equation}
The sum in Eq. (\ref{6}) is extended over a complete set of
final states $\phi_f(z)$ with energy $\varepsilon_f$ ($E_F\le E_f\le E_0$), $E_F$
is the Fermi  energy, and $W(z,z';{\bf q}_\parallel,\omega)$ is the
two-dimensional Fourier transform of the screened interaction:
\begin{eqnarray}\label{7}
W(&&z,z';{\bf q}_\parallel,\omega)=v(z,z';{\bf q}_\parallel)+\int{\rm
d}z_1\int {\rm d}z_2\cr\cr
&&\times v(z,z_1;{\bf q}_\parallel)\,\chi(z_1,z_2;{\bf
q}_\parallel,\omega)\,v(z_2,z';{\bf q}_\parallel),
\end{eqnarray}
with $v(z,z';{\bf q}_\parallel)$ and $\chi(z,z';{\bf q}_\parallel,\omega)$
representing the bare Coulomb interaction,
\begin{equation}\label{9}
v(z,z';{\bf q}_\parallel)={2\pi\over q_\parallel}\,e^{-q_\parallel\,|z-z'|},
\end{equation}
and the density-response function, respectively.

Within RPA, $\chi(z,z';{\bf q}_\parallel,\omega)$ satisfies the integral
equation 
\begin{eqnarray}\label{8}
\chi(&&z,z';{\bf q}_\parallel,\omega)=\chi_0(z,z',{\bf
q}_\parallel,\omega)+\int{\rm d}z_1\int{\rm d}z_2\cr\cr
&&\times\chi_0(z,z_1,{\bf q}_\parallel,\omega)\,v(z_1,z_2;{\bf q}_\parallel)
\,\chi(z_2,z',{\bf q}_\parallel,\omega),
\end{eqnarray}
where 
$\chi_0(z,z',{\bf q}_\parallel,\omega)$ represents the density-response
function of non-interacting electrons,\cite{Eguiluz}
\begin{eqnarray}\label{chi0}
\chi_0(&&z,z',{\bf q}_\parallel,\omega)=\sum_{l,l'}\phi_l(z)\phi_{l'}^*(z)
\phi_{l'}(z')\phi_l^*(z')\cr\cr
&&\times\int{{\rm d}{\bf
k}_\parallel\over(2\pi)^2}\,{\theta(E_F-E_l)-\theta(E_F-E_{l'})\over
E_l-E_{l'}+(\omega+{\rm i}\eta)},
\end{eqnarray}
with
\begin{equation}\label{4}
E_l=\varepsilon_l+{{\bf k}_\parallel^2\over 2}
\end{equation}
and
\begin{equation}\label{4}
E_{l'}=\varepsilon_{l'}+{({\bf k}_\parallel+{\bf q}_\parallel)^2\over
2},
\end{equation}
and $\theta(x)$ being the Heaviside step function.

In the so-called GW$\Gamma$ approximation, which includes XC effects not
present in the GW-RPA, the electron self-energy is of the GW form, i. e.,
it is given by Eq. (\ref{eq13}), but with an effective screened interaction
\begin{eqnarray}\label{eq7}
W(&&z, z';{\bf q}_\parallel,\omega)=v(z,z';{\bf q}_\parallel)+\int{\rm d}z_1\int
{\rm d}z_2\cr\cr
&&\times\left[v(z_1,z_2;{\bf q}_\parallel)+K_{xc}(z_1,z_2;{\bf
q}_\parallel,\omega)\right]\cr\cr &&\times\chi(z_1,z_2;{\bf
q}_\parallel,\omega)\,v(z_2,z';{\bf q}_\parallel),
\end{eqnarray}
where
\begin{eqnarray}\label{eq8}
\chi(&&z,z';{\bf q}_\parallel,\omega)=\chi_0(z,z',{\bf
q}_\parallel,\omega)+\int{\rm d}z_1\int{\rm d}z_2\cr\cr
&&\times\chi_0(z,z_1,{\bf q}_\parallel,\omega)\,
\left[v(z,z_1;{\bf q}_\parallel)\right.\cr\cr
&&\left.+K_{xc}(z_1,z_2;{\bf
q}_\parallel,\omega)\right]\,\chi(z_2,z',{\bf q}_\parallel,\omega).
\end{eqnarray}
Here, $K_{xc}(z,z';{\bf q}_\parallel,\omega)$ represents the two-dimensional
Fourier transform of the XC kernel $K_{xc}({\bf r},{\bf r}';\omega)$, which
accounts through Eqs. (\ref{eq7}) and (\ref{eq8}) for the reduction in the
electron-electron interaction due to the existence of short-range XC effects
associated to the probe electron and to screening electrons, respectively. In
the static limit ($\omega\to 0$), density-functional theory (DFT)\cite{DFT1}
shows that\cite{DFT2}
\begin{equation}\label{eq31}
K_{xc}({\bf r},{\bf r}';\omega\to 0)=\left[{\delta^2E_{xc}[n]\over\delta n({\bf
r})\delta n({\bf r}')}\right]_{n_0({\bf r})},
\end{equation}
where $E_{xc}[n]$ represents the XC energy functional and $n_0({\bf r})$ is
the actual density of the electron system. In the local-density approximation
(LDA), the XC kernel is approximated by a contact $\delta$ function, and one
finds
\begin{equation}\label{eq31}
K_{xc}(z,z';{\bf q}_\parallel,\omega\to 0)=\left[{{\rm d}^2E_{xc}(n)\over{\rm
d}n^2}\right]_{n_0(z)}\delta(z-z'),
\end{equation}
where $E_{xc}[n]$ now represents the XC energy of a homogeneous electron gas of
density $n$.\cite{Perdew} Introduction of this static XC kernel into Eq.
(\ref{eq8}) represents an adiabatic extension of the LDA to finite frequencies,
and yields the so-called time-dependent LDA (TDLDA).\cite{TDLDA}

The single-particle wave functions $\phi_i(z)$ entering Eqs. (\ref{5}),
(\ref{6}) and (\ref{chi0}) are simply eigenfunctions of a one-electron
hamiltonian. In the particular case of the RPA, $\phi_i(z)$ are self-consistent
eigenfunctions of the one-electron Hartree hamiltonian, and within TDLDA they
are obtained by solving the Kohn-Sham equation of DFT with use of the LDA XC
potential,
\begin{equation}\label{eq32}
V_{xc}(z)=\left[{{\rm d}E_{xc}(n)\over{\rm d}n}\right]_{n_0(z)}.
\end{equation}
Neither the Hartree self-consistent eigenfunctions nor the LDA wave functions
produce the correct image-like asymptotic potential behaviour on the vacuum
side of the surface.

For a realistic description of the metal surface, we
solve for $\phi_i(z)$ a single-particle time-independent Schr\"odinger
equation with the one-dimensional model potential suggested in
Ref.\onlinecite{Chulkov0}, which approaches, far outside the surface, the
classical image potential. This one-dimensional potential has four
adjustable parameters, which are chosen so as to reproduce the width and
position of the energy gap at the $\bar\Gamma$ point (${\bf k}_\parallel=0$)
and, also, the binding energies of both the $n=0$ crystal-induced surface
state at $\bar\Gamma$ and the first ($n=1$) image-potential induced state.
Probability amplitudes of the $n=1$ image state on Li(110) and Cu(100), as
obtained with use of this model potential, have been found to be in excellent
agreement with first principles pseudopotential\cite{Chulkov0} and
all-electron\cite{Hulbert} calculations, respectively.  

\section{Results and discussion}

Input of our calculation of image-state decay rates, as given by Eq.
(\ref{5}), are the image-state wave function $\phi_0(z)$ and the image-electron
self-energy $\Sigma(z,z';{\bf k_\parallel},E_0)$. We use films of up to 50
layers of atoms and 80 interlayer-spacing vacuum intervals, thereby ensuring
that finite-slab effects are negligible. Image-state wave functions are taken
to be eigenfunctions of the one-dimensional model hamiltonian described above
(MP). For the evaluation of the image-electron self-energy, we use in Eq.
(\ref{6}) either the MP wave functions or the self-consistent {\it jellium} LDA
eigenfunctions of the one-electron Kohn-Sham hamiltonian without (J) and with
(JG) the restriction that only final states with energy $\varepsilon_f$ lying
below the projected band gap are allowed. The screened interaction entering Eq.
(\ref{6}) is evaluated either within the specular-reflexion model (SRM) of
Ritchie and Marusak,\cite{SRM} with use of the approximate surface response
function of Persson and Zaremba (PZ),\cite{Zaremba} or within the
self-consistent approaches of Eqs. (\ref{7}) and (\ref{eq7}) with the
single-particle eigenstates entering Eq. (\ref{chi0}) being MP wave functions. 

Probability densities, $|\phi_0(z)|^2$, for the $n=1$ image state on the (100)
and (111) surfaces of Cu, as obtained with use of the one-dimensional
model potential of Ref.\onlinecite{Chulkov0}, are represented in Fig. 1. In the
case of Cu(100) (dotted line), the $n=1$ probability-density has a maximum at
$3.8\,\AA$ outside the crystal edge ($z=0$), which we choose to be located half
a lattice spacing beyond the last atomic layer, and the penetration into the
bulk ($z<0$) crystal is found to be of $5\%$. On the (100) surface of Cu, the
$n=1$ image state is close to the center of the projected band
gap,\cite{Fauster} which results in a very small penetration into the bulk.
However, the $n=1$ image state on Cu(111) (solid line) is located right at the
top of the gap,\cite{Fauster} the solution in the bulk is an
$s$-like wave function with the matching at the surface occurring at minimum amplitude, the
maximum of the probability-density in the vacuum is closer from the
surface, at $2.3\,\AA$, than in the case of Cu(100), and the penetration into
the bulk is found to be of $22\%$. The first
image state on Cu(100) and Cu(111) has binding energies of $0.57$ and
$0.82\,{\rm eV}$, respectively. 

Fig. 2 shows full GW-RPA calculations of the imaginary part of the $n=1$
image-electron self-energy ${\rm Im}\left[-\Sigma(z,z';{\bf
k_\parallel}=0,E_0)\right]$ in the vicinity of the (100) and
(111) surfaces of Cu, as obtained from Eq. (\ref{6}) with use of
MP wave functions both in Eq. (\ref{6}) and Eq. (\ref{chi0}) and with all
effective masses set equal to the free-electron mass. The imaginary part of the
electron self-energy is represented in this figure as a function of $z$ and for
a fixed value of $z'$. In the top panel,
$z'$ is fixed at about three atomic layers ($z'\sim-5\,\AA$) within the bulk,
showing that ${\rm Im}(-\Sigma)$ has a maximum at $z=z'$, as in the case of a
homogeneous electron gas. When $z'$ is fixed at the crystal edge ($z'\sim0$),
as shown in the middle panel of Fig. 2, we find that ${\rm Im}(-\Sigma)$ is
still maximum at $z=z'$, but the magnitude of this maximum now being enhanced.
The bottom panel of Fig. 2 corresponds to $z'$ being fixed at about three atomic
layers ($z'\sim 5\,\AA$) from the surface into the vacuum. In this case the
maximum magnitude of ${\rm Im}(-\Sigma)$ occurs at $z\sim 0$ rather than for
$z=z'$, showing a highly nonlocal behaviour of the imaginary part of the electron
self-energy at the surface. This nonlocality of the cusp of
${\rm Im}(-\Sigma)$ was also shown by Deisz {\it et al},\cite{Deisz} within a
free-electron (jellium) description of the surface.

The magnitude of the maximum of ${\rm Im}(-\Sigma)$ is plotted in Fig. 3, as a
function of $z'$, showing that it is an oscillating function of $z'$ within
the bulk and reaches its highest value near the surface. The oscillatory
behaviour within the bulk is dictated by the periodicity of final-state
wave functions $\phi_f(z)$ in the crystal, and the highest value near the
crystal edge is the result of electron-hole pair
creation taking place mainly in the vicinity of the surface. We note that
the magnitude of ${\rm Im}(-\Sigma)$ near the surface is larger for
Cu(111) than for Cu(100), although the band gap on
Cu(111) extends below the Fermi level, thus the available phase space on this
surface becoming highly restricted. However, while the crystal-induced
$n=0$ surface state on Cu(100) is located outside the projected band gap and
represents, therefore, a so-called surface resonance, the $n=0$ surface state on
Cu(111) is located within the band gap and
provides an important channel for the decay of image states. As a result, the
imaginary part of the image-electron self-energy near the (111) surface of
Cu is largely enhanced. This is illustrated in Fig. 4, where contributions to the
magnitude of the maximum ${\rm Im}(-\Sigma)$ of the $n=1$ image state on Cu(111)
have been plotted separately, according to whether transitions to bulk states
(dotted line) or to the $n=0$ surface state (dashed line) occur.

Now we focus on the evaluation of image-state lifetimes, and we set the wave vector of the image electron parallel to
the surface, ${\bf k}_\parallel$, equal to zero. Coupling of the image state
with the crystal occurs through the penetration of the image-state wave
function into the solid and, also, through the evanescent tails of bulk states
outside the crystal. Accordingly, we have calculated separately the various
contributions to the damping rate by confining the integral in Eq. (\ref{5}) to
either bulk ($z<0$) or vacuum ($z>0$) coordinates:
\begin{equation}
\tau^{-1}=\tau^{-1}_{bulk}+\tau^{-1}_{vac}+\tau^{-1}_{inter},
\end{equation}
where $\tau^{-1}_{bulk}$, $\tau^{-1}_{vac}$ and $\tau^{-1}_{inter}$ represent
bulk, vacuum and interference contributions, respectively. The results of our
calculations for the decay rate of the $n=1$ image state on the (100) and (111)
surfaces of Cu are presented in Tables I, II, and III.

In order to investigate the impact of band structure effects on the damping
rate of image states, we present in Table I the results of our full
GW-RPA calculations for the decay rate of the $n=1$ image state on Cu(100), as
obtained from Eq. (\ref{5}) with use of either J, JG or MP wave functions in Eq.
(\ref{6}), with use of MP wave fuctions in Eq. (\ref{chi0}), and with all
effective masses set equal to the free-electron mass. For
$q_\parallel>\sqrt{2(E_0-E_g)}$ ($E_g$ represents the bottom of the projected
band gap) all final states with energy
$E_f<E_0$ lie below the gap, thereby bulk and interference contributions to the
decay rate remaining nearly unaffected by this restriction. However, as the
coupling of the image state with the crystal occurring through the tails of bulk
states outside the crystal is expected to be dominated by vertical transitions
(${\bf q}_\parallel\sim 0$), the vacuum contribution to the decay rate becomes
noticeably smaller as final states lying within the projected band gap are not
allowed, and this restriction results in a total decay rate that is smaller by a
factor of $\sim 3$. However, the decay rate is nearly insensitive to the actual
choice of the one-particle wave functions entering Eq. (\ref{6}), as long as
only final states lying below the projected band gap are included in the
{\it jellium} calculation. We have also performed calculations with use of either
J or JG wave functions in Eq. (\ref{chi0}), and have found that the decay rate
is insensitive to the actual choice of the one-particle wave functions
entering Eq. (\ref{chi0}). In the case of the (111) surface of Cu, the bottom of
the projected band gap is located just below the Fermi level, and both the
existence of the $n=0$ surface state, not present within a jellium model of the
surface, and the restricted available phase space play a key role in the
determination of the $n=1$ image-state decay rate.

In Tables II and III we present the results of our calculations for the decay
rate of the $n=1$ image state on Cu(100) and (111), respectively, as obtained
from Eq. (\ref{5}) with use of MP wave functions in both Eqs. (\ref{6}) and
(\ref{chi0}). First of all, we set all effective masses equal to the
free-electron mass, and focus on the role that an accurate description of the
screened interaction plays in the coupling of image states with the solid.
Hence, we use three different models for the evaluation of $W(z,z';{\bf
q_\parallel},\omega)$. First, the specular-reflexion model (SRM) of Ritchie and
Marusak\cite{SRM} has been considered, thereby assuming that bulk electrons are
specularly reflected at the surface with no interference between ingoing and
outgoing waves. Secondly, for the vacuum contribution to the decay rate ($z>0$,
$z'>0$) the surface response function suggested by Persson and
Zaremba\cite{Zaremba} (PZ) has been used. Finally, the screened interaction has
been evaluated from Eq. (\ref{7}) by solving the RPA integral equation (Eq.
(\ref{8})) for the density-response function (GW-RPA).

We note that simplified jellium models (SRM and PZ) for the evaluation of the
screened interaction yield unrealistic results for the image-state lifetime.
Bulk contributions to the linewidth are approximately well described within the
specular reflexion model, small differences resulting from an approximate
description, within this model, of the so-called {\it bregenzung} effects.
However, as within this model quantum-mechanical details of the surface are
ignored, it fails to describe both vacuum and interference contributions to the
decay rate. These quantum-mechanical details of the surface are approximately
taken into account within the PZ jellium model, thereby
resulting in a better approximation for the vacuum contribution to the decay
rate than the SRM, but within the PZ model one neglects the coupling of the
image state with the crystal that occurs through the penetration of the
image-state wave function into the solid. Discrepancies between vacuum
contributions obtained within this model and our more realistic full RPA
calculations appear as a result of the jellium model of Ref.\onlinecite{Zaremba}
being accurate only for $q_\parallel/q_F$ and
$\omega/E_F<<1$ ($q_F$ is the Fermi momentum, i. e., $E_F=q_F^2/2$).

Now we look at the impact of short-range XC effects, which are well-known to
reduce electron-electron interactions both between the image electron and the
electron gas and between screening electrons themselves. Hence, we still set all
effective masses equal to the free-electron mass, we introduce MP wave
functions into Eqs. (\ref{6}) and (\ref{chi0}), and compare the GW-RPA
calculations described above with the results that we obtain from either
Eq. (\ref{7}) (GW) or Eq. (\ref{eq7}) (GW$\Gamma$) by using the density-response
function of Eq. (\ref{eq8}) with the LDA XC kernel of Eq. (\ref{eq31}) (TDLDA).
An inspection of the results presented in Tables I and II indicates that the
existence of XC effects between screening electrons enhances the decay
probability of image states. Nevertheless, this enhancement is more than
compensated by the large reduction of the decay rate produced by the presence of
a XC hole around the image-state electron. Consequently, GW-RPA calculations
produce decay rates that are larger than their more realistic GW$\Gamma$-TDLDA
counterparts by no more than $\sim 5\%$ in both (100) and (111) surfaces of
Cu.

Finally, we account for potential variation parallel to the surface through
the introduction of a realistic effective mass. The dispersion $E_0({\bf
k}_\parallel)$ of image states has been determined experimentally
with the use of inverse photoemission techniques at off-normal
emission,\cite{Fauster} showing that the effective mass of image states in both
Cu(100) and Cu(111) are $\sim 1$, i. e., the free electron mass. Measurements
of the dispersion of the $n=0$ surface resonance/state on the (100) and (111)
surfaces of Cu have yielded effective masses of $0.50$ and $0.42$,
respectively,\cite{Kevan,Goldmann,Hulbert2} and for bulk states entering Eq.
(\ref{6}) we have chosen to increase the effective mass from our computed
value of $m_f=0.40$ and $0.22$ at the bottom of the gap in Cu(100) and Cu(111),
respectively, to $m_f=1$ at the bottom of the valence band. As the effective mass
of all available final states is smaller than the free-electron mass, the
$n=1$ image-state decay rate of both Cu(100) and Cu(111) is found to be
about $20-25\%$ smaller than in the case of free-electron behaviour along the
surface ($m_f=1$). On the one hand,
there is the effect of the decrease of the available phase space, which is easily
found to scale as
$\sqrt{m_f}$. On the other hand, as the effective mass decreases the decay from
the image state occurs, for a given energy transfer, through smaller parallel
momentum transfer, which may result in either enlarged or diminished screened
interactions, depending on the magnitude of momentum and energy transfers. Our
results also indicate that GW-RPA and GW$\Gamma$-TDLDA calculations nearly
coincide, as in the case of free-electron behaviour along the surface, which is a
consequence of the competition between XC effects associated to the image-state
electron and to screening electrons themselves. GW$\Gamma$-TDLDA
calculations, as obtained with use of realistic effective masses for the
description of final-state wave functions, yield decay rates of the $n=1$ image
state on Cu(100) and Cu(111) of
$17$ and $28.5\,{\rm meV}$, respectively, in excellent agreement with the
experimentally determined lifetimes\cite{note1} of $40\pm 6$\cite{Hofer1,Hofer2}
and $22\pm 3\,{\rm fs}$.\cite{Hertel3}

With the aim of investigating the role that the various available bulk and
surface states play in the decay of image-potential states, now we focus on our
full GW-RPA calculation of the $n=1$ image-state decay rate. Fig. 5 exhibits
$\tau^{-1}_f$ separate contributions to $\tau^{-1}$ from all final
states lying below the projected band gap in Cu(100) (curves with circles)
and Cu(111) (curves with squares), as obtained with the final-state effective
mass set equal to the free-electron mass (solid lines) and with use of realistic
values for $m_f$ (dashed lines). In the case of Cu(111), there is still a large
contribution to $\tau^{-1}$ from the decay of the $n=1$
image state into the crystal-induced $n=0$ surface state, lying within
the projected band gap, which approximately represents $40\%$ of the
total decay rate. Figs. 5(a), (b), (c) and (d) exhibit bulk, vacuum,
interference, and total contributions to $\tau^{-1}_f$. As the $n=1$ image-state
wave-function overlap with the bulk is larger in Cu(111) than in the case of
Cu(100), bulk contributions to $\tau^{-1}_f$ decay rates are much larger for
Cu(111) than for Cu(100), as illustrated in Fig. 5(a). However, the large
bulk-state overlap in Cu(111) is partially counterbalanced by the band gap
extending on the (111) surface of Cu below the Fermi level. On the other hand,
vacuum and interference contributions to the decay rate in both surfaces of Cu
are comparable in magnitude and opposite in sign, yielding total decay
rates that differ little from the bulk contribution. We also note that the total
decay rate in Cu(100) ($22$ and $17.5\,{\rm meV}$, with $m_f=1$ and $m_f\neq 1$,
respectively) nearly coincides with the
$\sum_f\tau^{-1}_f$ contribution from all bulk states in Cu(111) ($21$ and
$17\,{\rm meV}$). Hence, differences between total decay rates in Cu(100) and
Cu(111) appear as a consequence of the presence of the $n=0$ surface state on
Cu(111), which provides a key channel for the decay mechanism.

Decay rates of image-potential states on Cu(100) with quantum
number $n\le 3$ are presented in Table IV, together with the experimentally
determined lifetimes reported in Refs.\onlinecite{Hofer1}
and\onlinecite{Hofer2}. As before, the wave vector of the image state parallel
to the surface, ${\bf k}$, has been set equal to zero, all the wave functions
entering Eqs. (\ref{5}), (\ref{6}) and (\ref{chi0}) have been chosen to be MP
wave functions, and the screened interaction has been evaluated within the
GW-RPA. Decay rates of the $n=2$ and $n=3$ image states have been split as
follows:
\begin{equation}
\tau^{-1}=\sum_f\tau^{-1}_f+\tau^{-1}_{n=1}
\end{equation}
and
\begin{equation}
\tau^{-1}=\sum_f\tau^{-1}_f+\tau^{-1}_{n=1}+\tau^{-1}_{n=2},
\end{equation}
respectively, where $\sum_f\tau^{-1}_f$ represents, as in the case of the $n=1$
image state, the decay rate from transitions to all available states lying
below the projected band gap, and $\tau^{-1}_n$ represents the contribution
from decay into the lower lying $n$ image state. We observe that lower lying
image states can give noticeable contributions to the decay rate of excited,
i. e., $n=2$ and $n=3$ image states. Decay of these states into the
$n=1$ image state results in linewidths that represent $\sim 10\%$ of the total
linewidth. Decay of the $n=3$ image state into the $n=2$ lower lying state
results in a linewidth that represents $\sim 3\%$ of the total linewidth. We
also observe that both our calculated and the experimentally determined
lifetimes of image states in Cu(100) increase with $n$ as $\sim1/n^3$, in
agreement with previous theoretical predictions.\cite{Echenique1} Discrepancies
between our calculated inelastic lifetimes of excited image states ($n>1$) and
experimental measurements may be attributed to scattering with phonons, which
occurs on a time scale ($\sim 1\,ps$) that is comparable for these states to the
electron-electron relaxation time. 

\section{Summary}

We have reported extensive self-consistent calculations of the imaginary part of
the electron self-energy and the decay rate of image states on the (100) and
(111) surfaces of Cu, by going beyond a free-electron description of the metal
surface. We have found that the imaginary part of the electron self-energy
outside the surface is highly nonlocal, and have found the magnitude of the
maximum of this quantity to reach its highest value near the
surface. We have presented the results of calculations of the lifetime of
the first three image states on Cu(100) and the first image state on
Cu(111), and have focused on the impact of band-structure and many-body
effects on these quantities. We have found that band-structure effects on the
evaluation of final-state wave functions may be approximately accounted through
introduction of the restriction that only final states lying below the
projected band gap are allowed, and the impact of the band structure through
the evaluation of the screened interaction has been found to be large. We have
shown that simplified jellium models for the electronic response yield
unrealistic results for the lifetime of image states, thereby a detailed
description of the screened interaction being of crucial importance in the
understanding of the origin of image-state lifetimes. We have evaluated the
screened interaction within three different self-consistent many-body schemes,
depending on whether XC effects are not included (GW-RPA) or they are included
through the introduction of a XC hole around screening electrons only (GW-TDLDA)
or around both the image-state electron and screening electrons
(GW$\Gamma$-TDLDA). We have reached the important conclusion that GW-RPA
calculations produce decay rates that are very close to GW$\Gamma$-TDLDA
calculations, which are obtained with full inclusion of XC effects. With the use
of either the free-electron mass or more realistic effective masses for all
final states, decay rates of the
$n=1$ image state on Cu(100) are found to be smaller than those of the $n=1$
image state on Cu(111). The large bulk-state overlap on Cu(111) is found to be
approximately counterbalanced by the band gap extending on the Cu(111) surface
below the Fermi level, and differences between decay rates on the (100) and
(111) surfaces of Cu are found to be mainly due to the large contribution, in the
case of Cu(111), from decay into the crystal-induced $n=0$ surface state
located within the projected band gap. The results we have
obtained for the $n=1$ image-state lifetime on both surfaces of Cu with use of
realistic effective masses, which are
$\sim 20\%$ below those obtained with $m_f=1$, are in excellent agreement with
experimentally determined decay times.

\section{Acknowledgments}
The authors gratefully acknowledge partial support by the University of
the Basque Country, the Basque Hezkuntza, Unibertsitate eta Ikerketa Saila, the
Spanish Ministerio de Educaci\'on y Cultura, and Iberdrola S. A..

\begin{table} \caption{GW-RPA decay rates, in linewidth units (${\rm
meV}$), of the $n=1$ image state on Cu(100), as obtained from Eqs. (\ref{5}),
(\ref{6}), (\ref{7}), and (\ref{8}). Effective masses have been set equal
to the free-electron mass ($m_f=1$). All wave functions entering Eqs.
(\ref{5}) and (\ref{chi0}) have been chosen to be MP wave functions, and the
final-state wave functions entering Eq. (\ref{6}) have been taken to be either
J, JG, or MP wave functions (see the text).}
\begin{tabular}{lcccccccc}
$\phi_f(z)$&Bulk&Vacuum&Inter.
&Total\\
\tableline
J&21&58&-12&67\\
JG&19.5&13&-11&21.5\\
MP&24&14&-16&22\\
\end{tabular}
\label{table1}  
\end{table}

\begin{table} \caption{Decay rates, in linewidth units (${\rm
meV}$), of the $n=1$ image state on Cu(100), together with the experimentally
determined decay rate of Refs.\protect\onlinecite{Hofer1}\protect\,
and\protect\onlinecite{Hofer2}\protect. All the wave functions entering Eqs.
(\ref{5}), (\ref{6}) and (\ref{chi0}) have been chosen to be MP wave functions.
Effective masses have been set equal to either the free-electron mass ($m_f=1$)
or to realistic values for all available final states ($m_f\neq 1$). Five
different models for the description of the screened interaction have been used:
the specular reflexion model (SRM),\protect\cite{SRM}\protect the model of
Persson and Zaremba (PZ),\protect\cite{Zaremba}\protect, and three
self-consistent many-body approaches, GW-RPA, GW-TDLDA, and GW$\Gamma$-TDLDA.}
\begin{tabular}{lcccccccc}
$m_f$&$W$&Bulk&Vacuum&Inter.
&Total&Exp.\\
\tableline
$=1$&SRM&18&3&-4&17\\
$=1$&PZ&&55&-&55\\
$=1$&GW-RPA&24&14&-16&22\\
$=1$&GW-TDLDA&26.5&14&-16&24.5\\
$=1$&GW$\Gamma$-TDLDA&23.5&14&-16&21.5\\
$\neq 1$&GW-RPA&7&11.5&-1&17.5\\
$\neq 1$&GW$\Gamma$-TDLDA&6.5&11.5&-1&17&16.5\\
\end{tabular}
\label{table1}  
\end{table}

\begin{table}
\caption{As in Table II, but for Cu (111) and together with the decay rate
experimentally determined for this surface
of copper\protect\cite{Hertel3}\protect at low temperature, $T=25\,K$.
Contributions to the linewidth from decay into bulk states lying below the
bottom of the projected band gap, thereby excluding the contribution from decay
into the $n=0$ intrinsic surface state, are displayed in parentheses.}
\begin{tabular}{lcccccccc}
$m_f$&$W$&Bulk&Vacuum&Inter.
&Total&Exp.\\
\tableline
$=1$&SRM&46(34)&12(1)&-22(-5)&36(30)\\
$=1$&PZ&-&57(2)&-&57(2)\\
$=1$&GW-RPA&44(28)&47(5)&-54(-12)&37(21)\\
$=1$&GW-TDLDA&43&42&-45&40\\
$=1$&GW$\Gamma$-TDLDA&43.5&47&-54&36.5\\
$\neq 1$&GW-RPA&32(24)&34(5)&-37(-12)&29(17)\\
$\neq 1$&GW$\Gamma$-TDLDA&30.5&35&-38&28.5&30\\
\end{tabular}
\label{table2}
\end{table}

\begin{table}
\caption{Calculated decay rates, in linewidth units (meV), and lifetimes,
in femtoseconds (fs), of the
$n\le 3$ image states on Cu(100), together with the experimentally determined
lifetimes of Refs.\protect\onlinecite{Hofer1}\,\protect and
\protect\onlinecite{Hofer2}\protect. In the case of
$n=2$ and $n=3$ contributions from decay into the lower lying image states are
also displayed. The screened interaction has been evaluated within the GW-RPA.
Model potential (MP) wave functions have been used for the description of all
single-particle states entering Eqs. (\ref{5}), (\ref{6}), and (\ref{chi0}).
The effective mass has been set equal to either the free-electron mass or to
the realistic values described in the text. The result of introducing into Eq.
(\ref{6}) a realistic effective mass for all available final states is
displayed in parentheses.} 
\begin{tabular}{lcccccccc}
&Linewidths&Lifetimes&Experiment\\
\tableline
$n=1$&22(17.5)&30(38)&$40\pm 6$\\
$n=2$&5(3.9)&132(168)&$120\pm 15$&&\\
$n=2$ to $n=1$&0.5(0.4)\\
$n=3$&1.8(1.4)&367(480)&$300\pm 20$&\\
$n=3$ to $n=2$&0.05(0.05)\\
$n=3$ to $n=1$&0.17(0.16)\\
\end{tabular}
\label{table6}
\end{table}

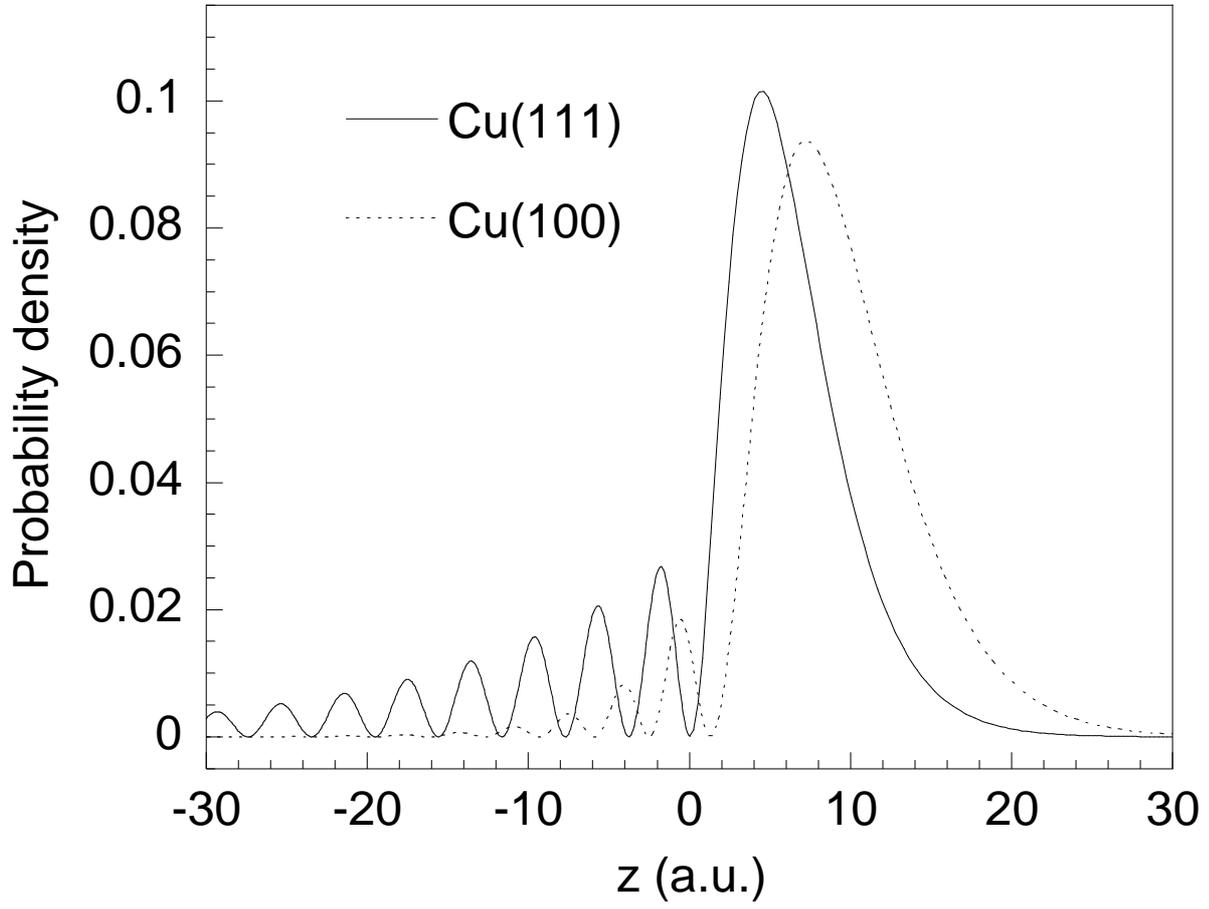
\begin{figure}
\caption{Probability density of the $n=1$ image state on Cu(100) (dotted line)
and Cu(111) (solid line), as obtained with use of the one-dimensional model
potential of Ref.\protect\onlinecite{Chulkov0}\protect. The crystal-edge ($z=0$)
has been chosen to be located half an interlayer spacing beyond the last atomic
layer. Full circles represent the atomic positions. In the case of Cu(111), the
matching at the surface occurs at minimum amplitude.}
\end{figure}

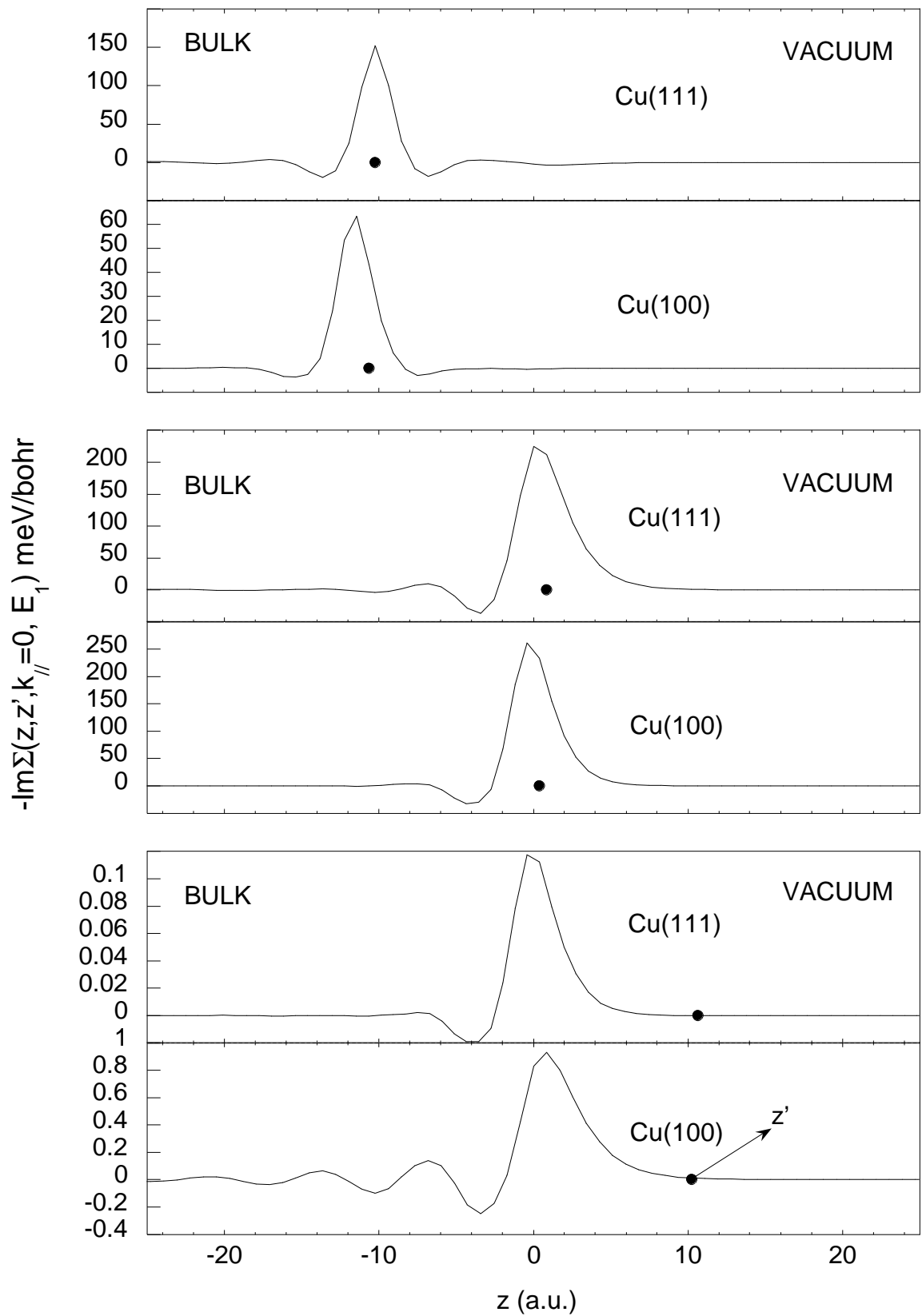
\begin{figure}
\caption{Imaginary part of the $n=1$ image-electron
GW-RPA self-energy
${\rm Im}\left[-\Sigma(z,z';{\bf k_\parallel}=0,E_0)\right]$, versus $z$, in the
vicinity of the (100) and (111) surfaces of Cu, as
obtained from Eq. (\ref{6}) with use of MP wave functions both in Eq.
(\protect\ref{6}\protect) and Eq. (\protect\ref{chi0}\protect) and with all
effective masses set equal to the free-electron mass. $z'$ is fixed at $-5$ (top
panel), $0$ (middle panel) and $5\,\AA$ (bottom panel).}
\end{figure}

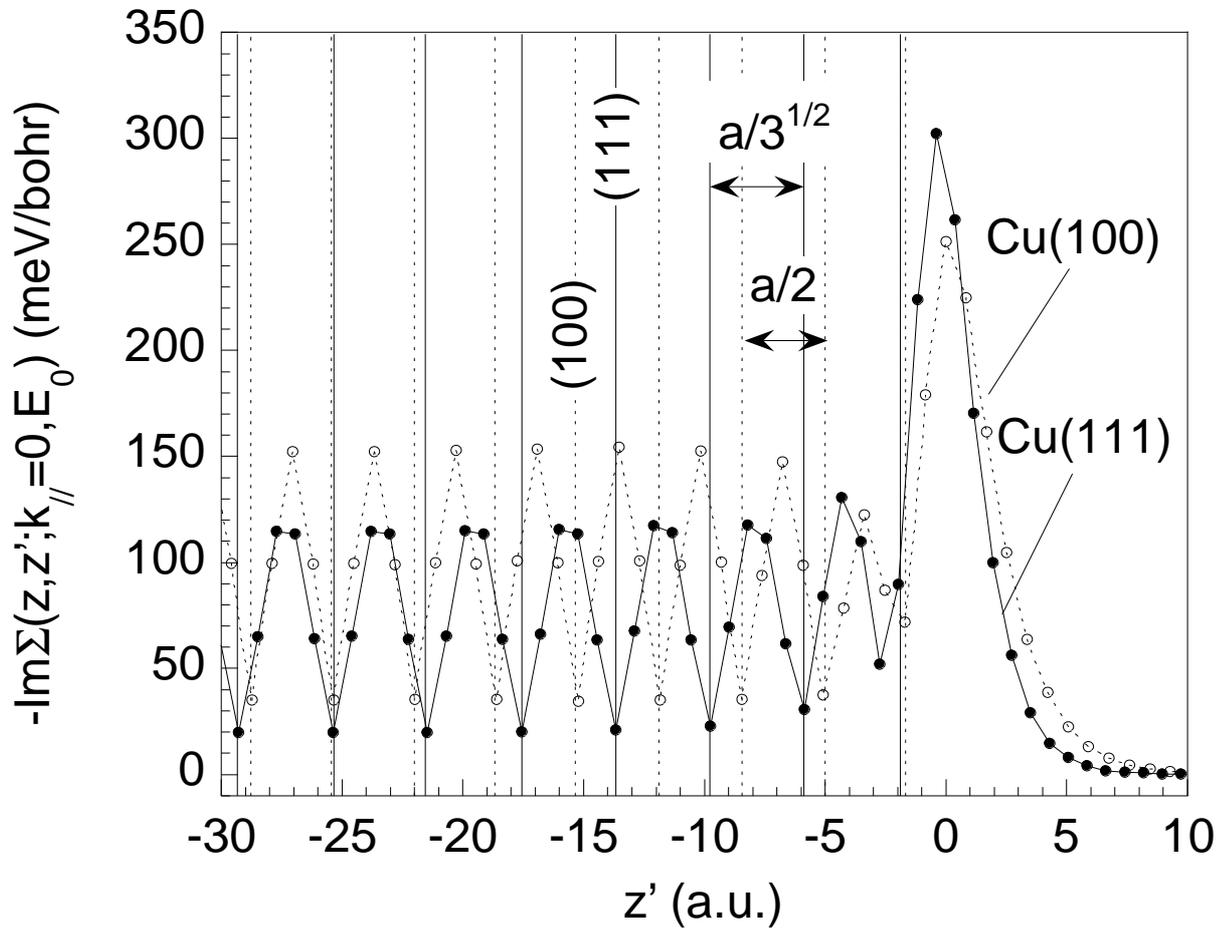
\begin{figure}
\caption{Maximum of the imaginary part of the $n=1$ image-electron GW-RPA
self-energy ${\rm Im}\left[-\Sigma(z,z';{\bf k_\parallel}=0,E_0)\right]$,
versus $z'$, in the vicinity of the  (100) (dotted lines) and (111) (solid
lines) surfaces of Cu, obtained as in Fig. 2. Vertical dotted and solid lines
represent the atomic positions along the (100) and (111) directions,
respectively.}
\end{figure}

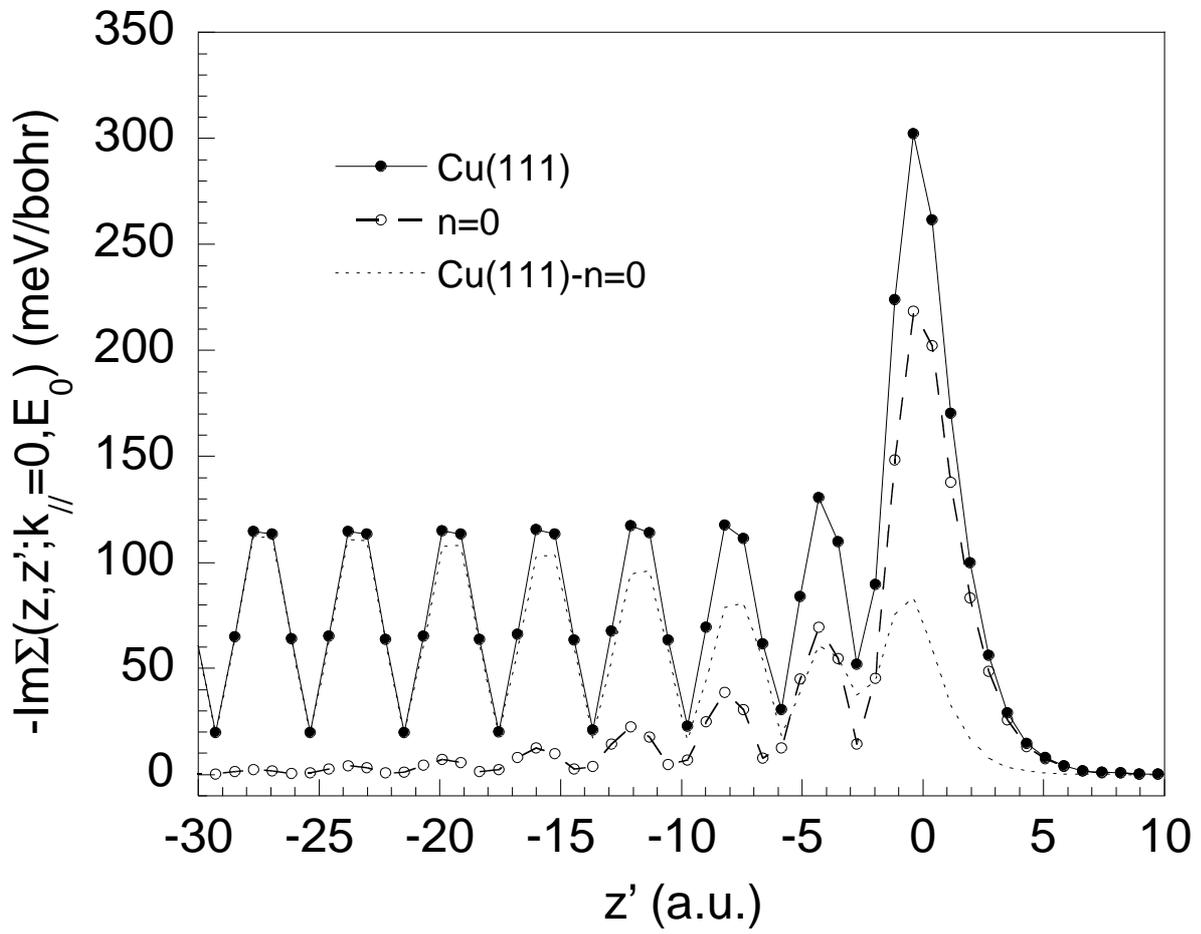
\begin{figure}
\caption{As in Fig. 4, for the separate contributions to the magnitude
of the maxiumum of ${\rm Im}\left[-\Sigma\right]$ of the $n=1$
image electron on Cu(111), according to wether transitions to bulk states
(dotted line) or to the $n=0$ surface state (dashed line) occur.}
\end{figure} 

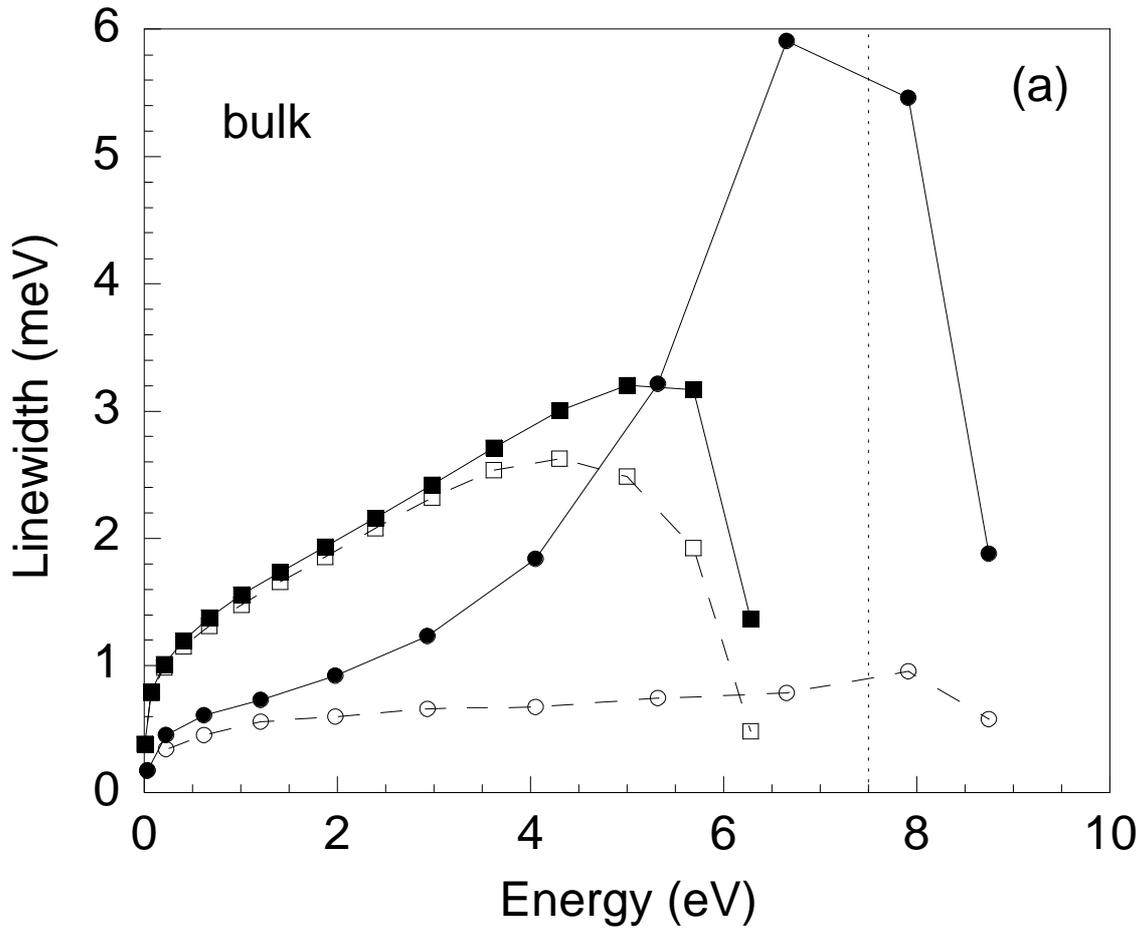
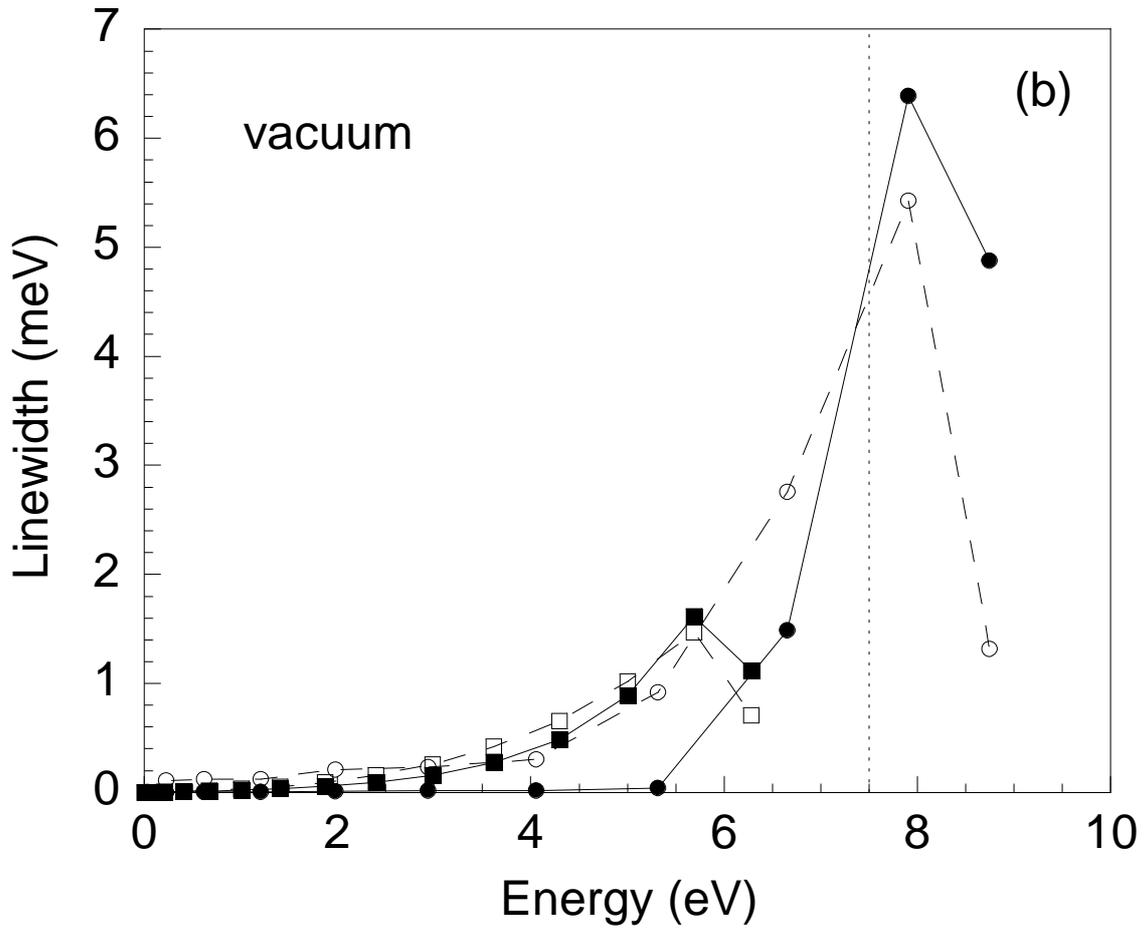
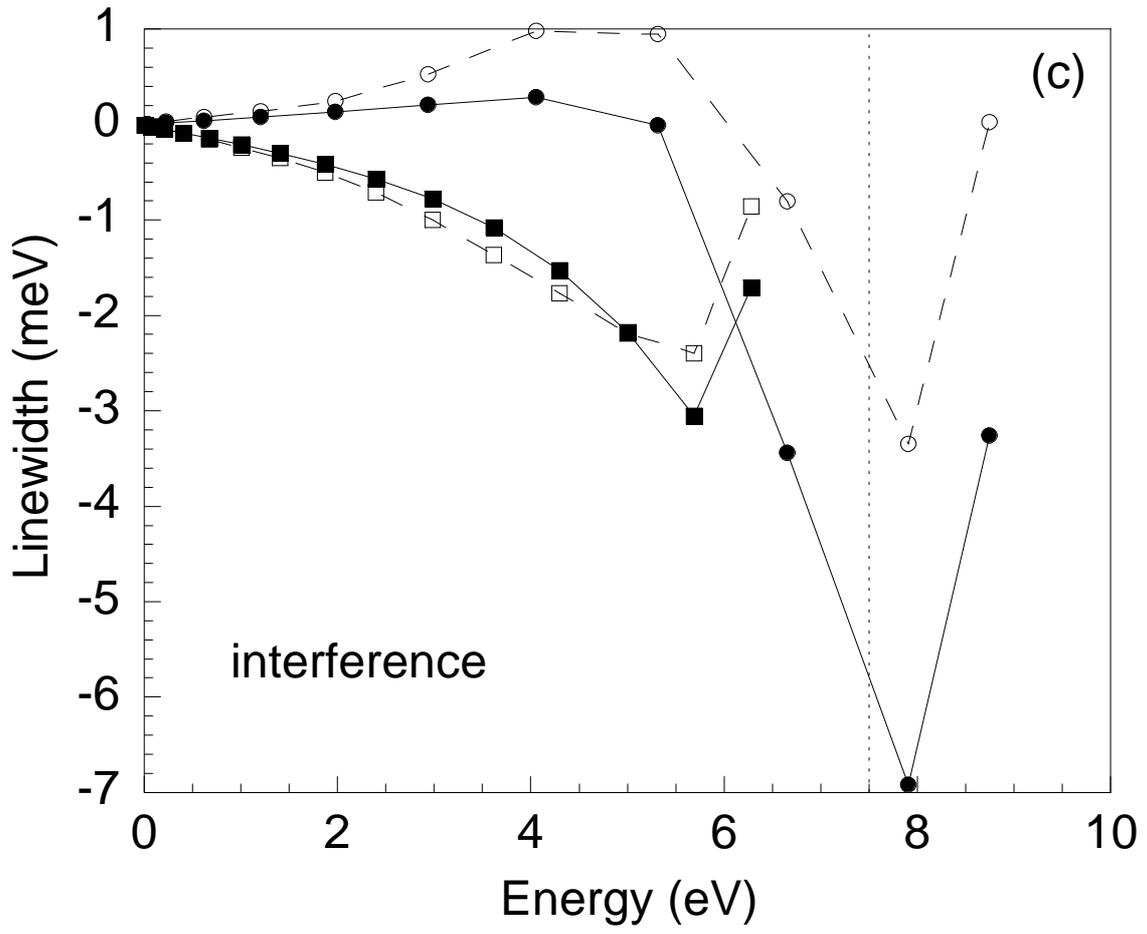
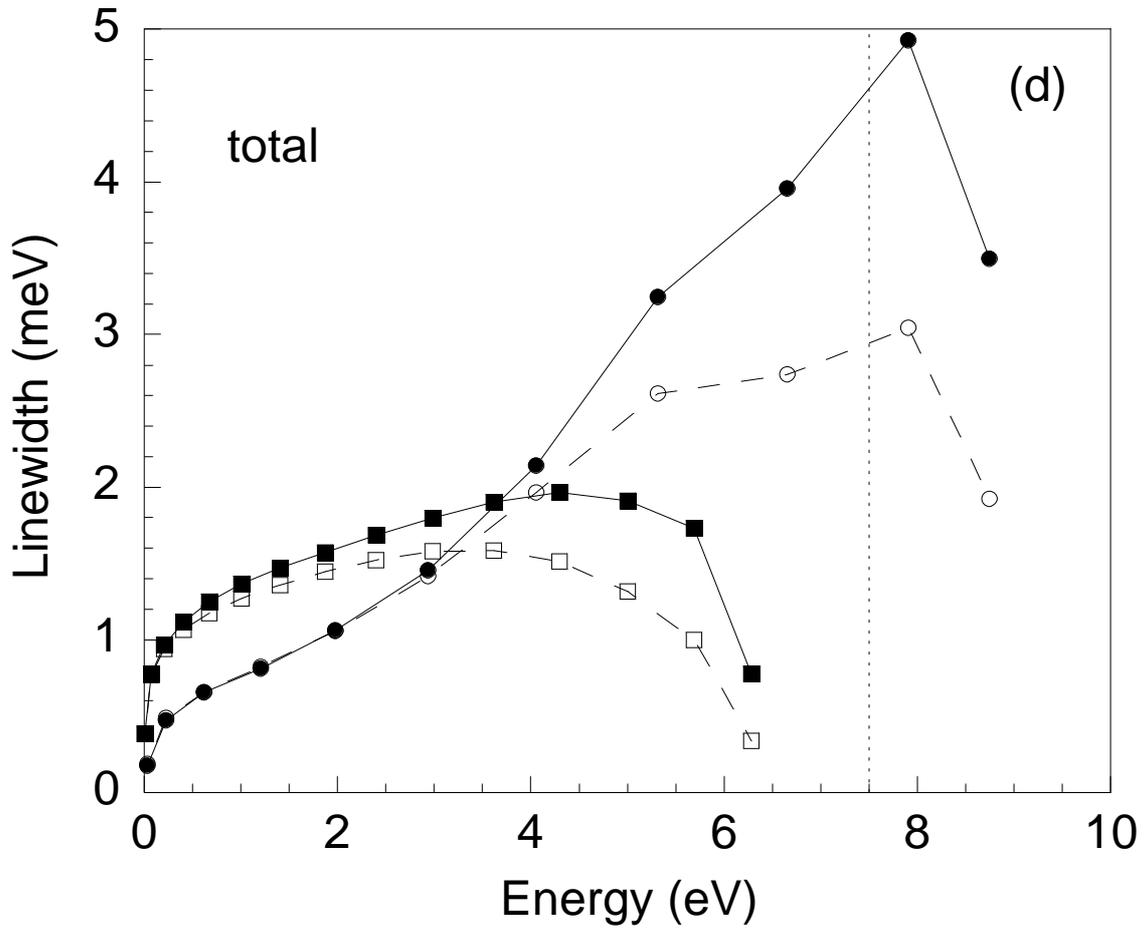
\begin{figure}
\caption{(a) Bulk, (b) vacuum, (c) interference, and (d) total contributions
to the GW-RPA damping rate $\tau^{-1}$ coming from the decay into the various $f$
available states lying below the projected band gap in Cu(100) (curves with
circles) and Cu(111) (curves with squares), as obtained with the final-state
effective mass set equal to the free-electron mass (solid lines) and with use
of realistic values of $m_f$ (dashed lines). Model potential (MP) wave functions
have been used for the description of all single-particle states entering Eqs.
(\ref{5}), (\ref{6}), and (\ref{chi0}). Vertical dotted lines represent the
Fermi level.}
\end{figure}


\begin{references}
\bibitem{note1} The decay rate of Eq. (\protect\ref{5}\protect) is related to the
lifetime, $\tau$, by $\tau^{-1}\tau=1\,{\rm a.u.}=658\,{\rm meV}\,{\rm fs}$.
\end{references}
\end{document}